\documentclass[preprint,showpacs,pra,superscriptaddress]{revtex4-1}
\usepackage[latin9]{inputenc}
\usepackage{amsmath}
\usepackage{amssymb}
\usepackage{graphicx}

\begin{document}

\title{Long-distance quantum key distribution with imperfect devices}

\author{Nicol\'o Lo Piparo}

\affiliation{School of Electronic and Electrical Engineering, University of Leeds,
Leeds, UK}

\author{Mohsen Razavi}

\email{m.razavi@leeds.ac.uk}

\affiliation{School of Electronic and Electrical Engineering, University of Leeds,
Leeds, UK}
\begin{abstract}
Quantum key distribution over probabilistic quantum repeaters is addressed.
We compare, under practical assumptions, two such schemes in terms
of their secure key generation rates per quantum memory. The two schemes
under investigation are the one proposed by Duan {\em et al.} in
{[}Nat. \textbf{414,} 413 (2001){]} and that of Sangouard {\em et
al.} in {[}Phys. Rev. A \textbf{76,} 050301 (2007){]}. We consider
various sources of imperfection in both protocols, such as nonzero
double-photon probabilities at the sources, dark counts in detectors,
and inefficiencies in the channel, photodetectors and memories. We
also consider memory decay and dephasing processes in our analysis.
For the latter system, we determine the maximum value of the double-photon
probability beyond which secret-key distillation is not possible.
We also find crossover distances for one nesting level to its subsequent one. We finally compare the two protocols
in terms of their achievable secret key generation rates at their
optimal settings. Our results specify regimes of operation where one system outperforms the other. 
\end{abstract}


\maketitle

\section{introduction}

\label{Sec:Intro} Despite all practical progress with quantum key
distribution (QKD) \cite{Wang:260kmQKD:2012,Tokyo_QKD,Townsend_QI_home_2011,Razavi_MA_QKD}, its implementation over long distances remains
to be a daunting task. In conventional QKD protocols such as BB84
\cite{BB84:1}, channel loss and detector noises set an upper bound
on the achievable security distance \cite{Lo:Decoy:2005}. In addition, the path loss results
in an exponential decay of the secret key generation rate with distance. Both these issues can, in principle, be overcome
if one implements entanglement-based QKD protocols \cite{PhysRevLett.67.661, BBM_92} over quantum repeater
systems \cite{PhysRevLett.81.5932,:1, PhysRevA.82.032304, Von_Loock_QKD_repeater_1}.
This approach, however, is not without its own challenges. Quantum
repeaters require quantum memory (QM) units that can interact with
light and can store their states for sufficiently long times. Moreover,
highly efficient quantum gates might be needed to perform two-qubit
operations on these QMs \cite{PhysRevLett.81.5932}. The latter issue
has been alleviated, to some extent, by introducing a novel technique
by Duan, Lukin, Cirac and Zoller (DLCZ) \cite{:1}, in which initial
entanglement distribution and swapping, thereafter, rely on probabilistic
linear-optic operations. Since its introduction, the DLCZ idea has
been extended and a number of new proposals have emerged \cite{jiang,PhysRevA.76.022329,PhysRevLett.98.190503,Phys_rev_A77, PhysRevA.76.050301, RevModPhys.83.33}.
Such probabilistic schemes for quantum repeaters particularly find
applications in QKD systems of mid-to-long distances, which makes
them worthy of analytical scrutiny. This papers compares DLCZ with
one of its favorite successors, \cite{PhysRevA.76.050301}, which relies
on single photon sources (termed SPS, hereafter). Using a general system-level approach, which encompasses many relevant physical
sources of imperfections in both systems, we provide a realistic account
of their performance in terms of their secret key generation rates
per logical memory used. This measure not only quantifies performance,
but it also accounts for possible costs of implementation.

The SPS protocol attempts to resolve one of the key drawbacks in the
original DLCZ protocol: multi-photon emissions. DLCZ uses atomic ensembles
as QMs, which lend themselves to multi-photon emissions. This leads
to obtaining not-fully-entangled states, hence resulting in lower
key rates when used for QKD. To tackle this issue, in the SPS protocol,
entanglement is distributed by ideally generating single photons,
which will either be stored in QMs, or directed toward a measurement
site. Whereas, in principle, the SPS protocol should not deal with
the multi-photon problem, in practice, it is challenging to build
on-demand single photon sources that do not produce any multi-photon
components. A fair comparison between the two systems is
only possible when one considers different sources of non-idealities
in both cases, as we will pursue in this paper.

The SPS protocol is one of the many proposed schemes for probabilistic
quantum repeaters. In \cite{RevModPhys.83.33}, authors provide a
review of all such schemes and compare them in terms of the average
time that it takes to generate entangled states, of a certain {\em
fidelity}, between two remote memories. Their conclusion is that
in the limit of highly efficient memories and detectors, the top three
protocol are the SPS protocol and two others that rely on entangled/two
photon sources \cite{Phys_rev_A77,PhysRevA.76.022329}. In more practical
regimes, however, the SPS protocol seems to have the best performance
per memory/mode used. In this paper, we therefore focus on the SPS
protocol, and will investigate, under practical assumptions, whether
the above conclusion remains valid in the context of QKD systems.

Our work is distinct from previous related work in its focusing on the performance of {\em QKD} systems over quantum repeaters. In \cite{RevModPhys.83.33}, authors have adopted the general measure of fidelity to find the average time of entanglement generation. Whereas their approach provides us with a general insight into some aspects of quantum repeater systems, it cannot
be directly applied to the case of QKD. In the latter, the performance is not only a function of the entanglement generation rate, but also the quantum bit error rate caused by using non-ideal entangled states. To include both these issues, here, we adopt the secret-key
generation rate per memory as the main figure of merit, by which we
can specify the optimal setting of the system and its performance
in different regimes of operation.

Another key feature of our work is to use a {\em normalized} figure
of merit to compare the DLCZ and SPS protocols. In practice, to obtain
a sufficiently large key rate in such probabilistic systems, one must
use multiple memories and/or modes in parallel. In order to account
for the cost of the system, in our analysis, we provide a normalized
key rate per memory and/or mode. We calculate the dependence of the
secret key generation rate on different system parameters when resolving
or non-resolving detectors are used. In particular, we find the optimal
values for relevant system parameters if loss, double-photon emissions
and dark counts are considered. Moreover, we account for the dephasing
and the decay of memories in our analysis, which, we believe, is unprecedented.

The paper is structured as follows. In Sec. II, we review the DLCZ
and the SPS protocols, their entanglement distribution and swapping
schemes, as well as their QKD measurements. In Sec. III, we present
our methodology for calculating the secret key generation rate for
the SPS protocol, followed by numerical results in Sec. IV. We draw
our conclusions in Sec. V.

\section{two probabilistic schemes for quantum repeaters}

In this section we will review two probabilistic schemes, namely,
DLCZ and SPS, for quantum repeaters. We describe the multiple-memory
setup for such systems and model relevant system components. 

\subsection{DLCZ entanglement-distribution scheme}

\begin{figure}
\includegraphics[width=8cm]{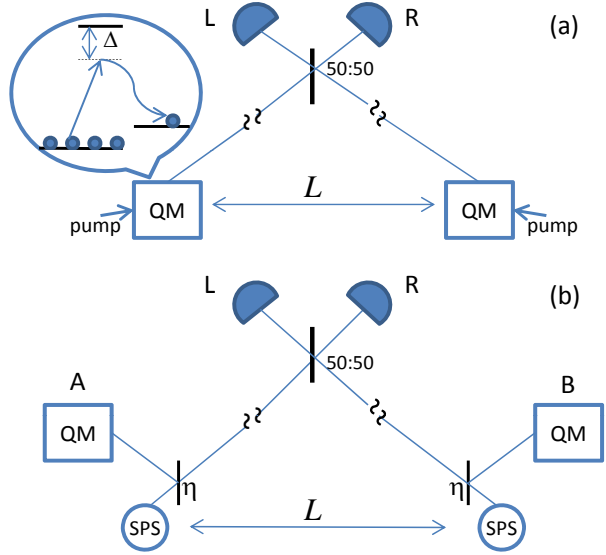}
\caption{{\footnotesize \label{fig:DLCZ and SPS} (Color online) Schematic diagram for entanglement
distribution between quantum memories (QMs) A and B for (a) the DLCZ
protocol and (b) the SPS protocol. In both cases, we assume QMs can
store multiple excitations. Sources, memories and detectors are represented
by circles, squares and half-circles, respectively. Vertical bars
denote beam splitters. In both protocols the detection of a single
photon ideally projects the two memories onto an entangled state.}}
\end{figure}

The DLCZ scheme works as follows; see Fig. \ref{fig:DLCZ and SPS}(a).
Ensemble memories $A$ and $B,$ at distance $L,$ are made of atoms
with $\Lambda$-level configurations. They are all initially in their
ground states. By coherently pumping these atoms, some of them may
undergo off-resonant Raman transitions that produce Stokes photons.
The resulting photons are sent toward a 50:50 beam splitter located
at distance $L/2$ between $A$ and $B.$ If, ideally, only one photon
has been produced in total at the ensembles, one and, at most, only
one of the detectors in Fig. \ref{fig:DLCZ and SPS}(a) clicks. In
such a case, the DLCZ protocol heralds $A$ and $B$ to be ideally
in one of the Bell state $\left|\psi_{\pm}\right\rangle _{AB}=\left(|10\rangle_{AB}\pm|01\rangle_{AB}\right)/\sqrt{2},$
where $|0\rangle_{J}$ is the ensemble ground state and $|1\rangle_{J}=S_{J}^{\dagger}|0\rangle_{J}$
is the symmetric collective excited state of ensemble $J=A,\, B,$
where $S_{J}^{\dagger}$ is the corresponding creation operator \cite{:1}.
An important feature of such collective excitations is that they can
be read out by converting their states into photonic states. 

The fundamental source of error in the DLCZ scheme is the multiple-excitation
effect, where more than one Stokes photon are produced \cite{PhysRevA.82.032304}.
If the probability of generating one Stokes photon is denoted by $p_{c}$,
there is a probability $p_{c}^{2}$ that each ensemble emits one photon.
If this happens, a click on one of the two detectors heralds entanglement
generation, whereas the memories are in the separable state $|11\rangle_{AB}$.

In practice, one has to find the right balance between the heralding
probability, which increases with $p_{c}$, and the quantum bit error
rate (QBER), which also increases with $p_{c}$. In \cite{PhysRevA.82.032304},
authors find the optimal value of $p_{c}$ that maximizes the secret
key generation rate in various scenarios when photon-number resolving
detectors (PNRDs) or non-resolving photon detectors (NRPDs) are used.
In this paper, we use their results in our comparative study.

\subsection{SPS entanglement-distribution scheme}

The SPS protocol, proposed in \cite{PhysRevA.76.050301}, aims at
reducing multi-photon errors and, in particular, terms of the form
$|11\rangle_{AB}$ by using single-photon sources. The architecture
of this scheme is presented in Fig.~\ref{fig:DLCZ and SPS}(b). The
two remote parties each have one single-photon source and one memory.
In the ideal scenario, each source produces exactly one photon on demand,
and these photons are sent through identical beam splitters with transmission
coefficients $\eta.$ It can be shown that the state shared by the
QMs after a single click on one of the detectors in Fig.~\ref{fig:DLCZ and SPS}(b) is given by \cite{PhysRevA.76.050301}
\begin{equation}
\eta|00\rangle_{AB}\langle00|+(1-\eta)\left|\psi_{\pm}\right\rangle _{AB}\left\langle \psi_{\pm}\right|,\label{Eq:SPSout}
\end{equation}
 which has our desired entangled state plus a vacuum component. The
latter, at the price of reducing the rate, can be selected out once
the above state is measured at later stages \cite{:1,PhysRevA.82.032304}.

In a practical setup, several sources of imperfection must be considered
in Fig.~\ref{fig:DLCZ and SPS}(b). First, most known techniques
for generating single photons suffer from multiple-photon emissions.
That includes single-photon sources that rely on parametric down-conversion \cite{PhysRevA.79.035801,numero2},
quasi-atomic structures such as quantum dots \cite{Qdot_source_1}, or the partial
memory-readout technique described in \cite{RevModPhys.83.33}. In
all cases, there is a nonzero probability to generate more than one
photon, which manifests itself in producing nonzero values for second-order
coherence functions \cite{PhysRevA.79.035801,numero2}. For practical
purposes, however, it is often sufficient to consider the effect of
two-photon states, as we will do, in this paper. It turns out that
this approximation is particularly valid for the systems of interest
in this paper. One should also consider non-idealities in QMs. In
our analysis, we account for reading and writing efficiencies of QMs,
as well as their decay and dephasing processes. We assume that QMs
can store multiple excitations.

Throughout the paper, we assume that both setups in Fig.~\ref{fig:DLCZ and SPS}
are symmetric and phase stabilized. Furthermore, all conditions required
for a proper quantum interference at 50-50 beam splitters are assumed
to be met. Recent experimental progress in QKD shows that it is indeed
possible to achieve these conditions \cite{Lo:MIQKD:2012,Rubenok:MIQKDexp:2012}.

\subsection{Entanglement swapping and QKD measurements}

\begin{figure}
\includegraphics[width=8cm]{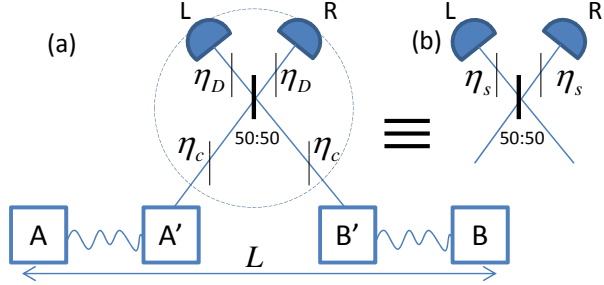}
\caption{{\footnotesize \label{fig:BSM} (Color online) (a) Entanglement connection between
two entangled links $A$-$A'$ and $B'$-$B$. The memories $A'$ and $B'$
are read out and the resulting photons are combined on a 50:50 beam
splitter. A click on one of the detectors projects $A$ and $B$ into
an entangled state. The retrieval efficiencies and quantum efficiencies
are represented by fictitious beam splitters with transmission coefficient
$\eta_{c}$ and $\eta_{D},$ respectively. (b) The equivalent butterfly
transformation to the measurement module, where $\eta_{s}=\eta_{c}\eta_{D}.$}}
\end{figure}

Figure \ref{fig:BSM}(a) shows the entanglement swapping setup for
the DLCZ and the SPS protocols. Entanglement is established between
QM pairs $AA'$ and $B'B$ using either of protocols. A partial Bell-state
measurement (BSM) on photons retrieved from the middle QMs $A'$ and
$B'$ is then followed, which upon success, leaves $A$ and $B$ entangled.
The BSM is effectively performed by a 50:50 beam splitter and single-photon
detectors. To include the effects of the atomic-to-photonic conversion
efficiency and the photodetectors' quantum efficiency, we introduce
two fictitious beam splitters with transmission coefficients $\eta_{c}$
and $\eta_{D},$ respectively. All photodetctors in Fig.~\ref{fig:BSM}
will then have unity quantum efficiencies. Note that the parameter
$\eta_{c}$ also includes the memory decay during the storage time.

Figure \ref{fig:BSM}(b) provides a simplified model for the measurement
module in Fig. \ref{fig:BSM}(a). The 50:50 beam splitter and the
two fictitious beam splitters in Fig. \ref{fig:BSM}(b) constitute
what we call a {\em butterfly} operation, which will be further
studied in Sec.~\ref{Sec:SPS} and Appendix \ref{Sec:Buterfly}.

\begin{figure}
\includegraphics[width=8.5cm]{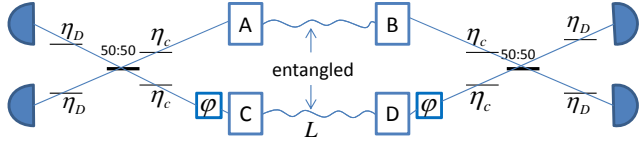}
\caption{{\footnotesize \label{fig:QKD-measurements} (Color online) QKD measurements on two
entangled pairs. Two pairs of memories, $A$-$B$ and $C$-$D$, each
share an entangled state. Memories are read out and the resulting
photons are combined at a beam splitter and then detected. Different
QKD measurements can be performed by choosing different phase shift
values, $\varphi$, of $0$ and $\pi/2$. }}
\end{figure}

Alice and Bob use two butterfly operations to generate a raw key bit,
as shown in Fig.~\ref{fig:QKD-measurements}. After generating entangled
pairs over a distance $L$, Alice and Bob retrieve the states of memories
and perform a QKD measurement on the resulting photons. They apply
a random relative phase shift, $\varphi,$ of either $0$ or ${\pi}/{2}$,
between their two fields. They will later, at the sifting stage, only
keep data points where the same phase value is used by both parties.
They then turn their sifted keys into a secure key by using privacy
amplification and error reconciliation techniques. Eavesdroppers can
be detected by following the BBM92 or the Ekert protocol \cite{PhysRevLett.68.3121,PhysRevLett.67.661}.

As mentioned in Sec.~\ref{Sec:Intro}, previous analyses only provide
the fidelity or the time required for a successful creation of an
entangled state \cite{PhysRevA.76.050301}. Instead, in Sec.~\ref{Sec:SPS},
we will calculate the secret key generation rate for the SPS scheme
and compare it with that of the DLCZ protocol reported in \cite{PhysRevA.82.032304}.

\subsection{Multiple-memory configuration}

\begin{figure}
\begin{centering}
\includegraphics[width=8.6cm]{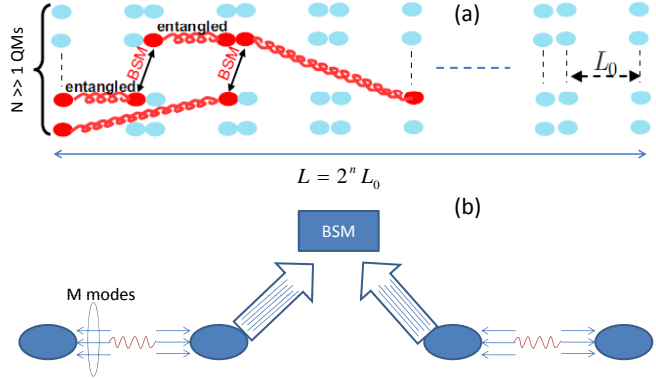} 
\par\end{centering}
\caption{{\footnotesize \label{fig: mult mem conf} (Color online) (a) A quantum repeater with
multiple quantum memories per node. At each round, we employ entanglement
distribution protocol to entangle any unentangled memory pairs over
shortest links. At any such cycle, we also match up entangled pairs
at different stations to perform Bell-state measurements (BSMs). (b)
A quantum repeater with multimode memories. In each round, we apply
our entanglement distribution scheme on all M modes, until one of
them becomes entangled. BSM will be followed as soon as entanglement
is established on both sides.}}
\end{figure}

In order to compare different quantum repeater setups, we consider
the multiple-memory configuration shown in Fig. \ref{fig: mult mem conf}(a)
along with the cyclic protocol described in \cite{PhysRevA.80.032301,Razavi_SPIE}.
In this protocol, in every cycle of duration $L_{0}/c,$ where $L_{0}$
is the length of the shortest segment in a quantum repeater, and $c$
is the speed of light in the channel, we try to entangle any unentangled
pairs of memories at distance $L_{0}.$ We assume our entanglement-distribution
protocol succeeds with probability $P_{S}\left(L_{0}\right)$. At
each cycle, we also perform as many BSMs as possible at the intermediate
nodes. The main requirement for such a protocol is that, at the stations
that we perform BSMs, we must be aware of establishment of entanglement
over links of length $l/2$ before extending it to $l$ (\textit{informed}
BSMs). We use the results of \cite{PhysRevA.80.032301} to calculate
the generation rate of entangled states {\em per memory} in the
limit of infinitely many memories. It is given by $R_{{\rm ent}}\left(L\right)=P_{S}\left(L/2^{n}\right)P_{M}^{\left(1\right)}P_{M}^{\left(2\right)}...P_{M}^{\left(n\right)}/\left(2L/c\right),$
where $P_{M}^{\left(i\right)},$ $i=1...n,$ is the BSM success probability
at nesting level $i$ for a quantum repeater with $n$ nesting levels.

We use the following procedure, in forthcoming sections, to find the
secret key generation rate of the setup in Fig. \ref{fig: mult mem conf}(a).
For each entanglement distribution scheme, we find $P_{S}\left(L_{0}\right)$
and relevant $P_{M}$ probabilities to derive $R_{{\rm ent}}\left(L\right).$
We then find the sifted key generation rate by multiplying $R_{{\rm ent}}\left(L\right)$
by the probability, $P_{{\rm click}},$ that an acceptable click pattern
occurs upon QKD measurements. Finally, the ratio between the number
of secure bits and the sifted key bits is calculated using the Shor-Preskill
lower bound \cite{PhysRevLett.85.441}. In the limit of an infinitely
long key, the secret key generation rate per logical memory is lower
bounded by 
\begin{equation}
R_{{\rm QKD}}\left(L\right)=\max(R_{{\rm ent}}\left(L\right)P_{{\rm click}}\left[1-2\, H\left(\epsilon_{Q}\right)\right],0),\label{eq:equazione 2}
\end{equation}
 where $\epsilon_{Q}$ denotes the QBER, and $H(p)=-p\log_{2}p-(1-p)\log_{2}(1-p),$
for $0\leq p\leq1.$

\subsection{Multimode-memory configuration }

Another way to speed up the entanglement generation rate is via using
multimode memories \cite{PhysRevLett.98.190503, Afzelius2009}. As can be seen in
Fig. \ref{fig: mult mem conf}(b), in this setup, we only use one
physical memory per node but each memory is capable of storing multiple
modes. In each round, we attempt to entangle memories at distance
$L_{0}$ by entangling, at least, one of the existing $M$ modes.
Once this occurs, we stop entanglement generation on that leg and
wait until a BSM can be performed. For readout, all modes must be
retrieved in order to perform BSMs or QKD measurements on particular
modes of interest. In effect, this scheme is similar to that of Fig.
\ref{fig: mult mem conf}(a), except that entanglement distribution
is not sequentially applied to unentangled modes. The success probability
for entanglement distribution between the two memories is, however,
$M$ times that of Fig. \ref{fig: mult mem conf}(a). One can show
that, the generation rate of entangled states per mode is approximately
given by $\left(\frac{2}{3}\right)^{n}R_{{\rm ent}}\left(L\right)$
\cite{Razavi_SPIE,RevModPhys.83.33}. 

In our forthcoming analysis, we only consider the case of Fig. \ref{fig: mult mem conf}(a),
but our results are extensible to the case of Fig. \ref{fig: mult mem conf}(b)
by accounting for the relevant prefactor.

\subsection{Memory decay and dephasing}

Quantum memories are expected to decay and dephase while storing quantum
states. In this paper, we model these two processes independently.
The decay process, with a time constant $T_{1}$, can be absorbed
in the retrieval efficiency of memories. If the retrieval efficiency
immediately after writing into the memory is given by $\eta_{0}$,
after a storage time $T$, the retrieval efficiency is given by $\eta_{c}=\eta_{0}\exp(-T/T_{1})$.
Different memories in the multiple-memory setup of Fig.~\ref{fig: mult mem conf}(a)
undergo different decay times. In our analysis, we consider the worst
case scenario where all memories have decayed for $T=L/c$, which
is only applicable to the far-end memories. Under this assumption,
$\eta_{c}$ can be treated as a constant at all stages of entanglement
swapping.

We model the memory dephasing via a dephasing channel, by which the
probability of dephasing after a period $T$ is given by $e_{d}=[1-\exp(-T/T_{2})]/2$.
In the context of the QKD protocol in Fig.~\ref{fig:QKD-measurements},
this phase error is equivalent to the misalignment error in a conventional
polarization-based BB84 protocol and has mostly the same effect. In
our analysis, we neglect the effect of dephasing at the middle stages,
and only consider its effect on the far-end memories used for the
QKD protocol. Again, for the multiple-memory setup of Fig.~\ref{fig: mult mem conf}(a),
the relevant storage time is given by $T=L/c$ \cite{PhysRevA.80.032301}.

\section{SPS secret key generation rate}

\label{Sec:SPS} In this section, the secure key generation rate for
the SPS scheme proposed in \cite{PhysRevA.76.050301} is calculated.
As was shown in section II, this scheme relies on simultaneous generation
of single photons in two remote sites. Most practical schemes for
the generation of single photons, however, suffer from the possibility
of multiple-photon emissions. To address this issue, in this section,
we consider non-ideal photon sources with nonzero probabilities for
two-photon emissions, and find the secret key generation rate in the
repeater and no-repeater cases.

Suppose our photon sources emit one photon with probability $1-p$
and two photons with probability $p.$ We, therefore, have the following
input density matrix for the initial state of $l$ and $r$ sources
in Fig. \ref{fig:Model of setup}(a) 
\begin{equation}
\rho_{lr}^{\left(in\right)}=\rho_{l}^{\left(in\right)}\otimes\rho_{r}^{\left(in\right)},\label{eq:initial density matrix}
\end{equation}
 where 
\begin{equation}
\begin{array}{cc}
\rho_{j}^{\left(in\right)}\equiv\left(1-p\right)|1\left\rangle _{jj}\left\langle 1|+p|2\left\rangle _{jj}\left\langle 2|\right.\right.\right.\right., & j=l,\, r.\end{array}\label{eq:input_state}
\end{equation}
As we will show later, in a practical regime of operation, $p\ll1;$
hence, in our following analysis, we neglect $O\left(p^{2}\right)$
terms corresponding to the simultaneous emission of two photons by
both sources.

\subsection{No-repeater case}

\begin{figure}
\includegraphics[width=8.6cm]{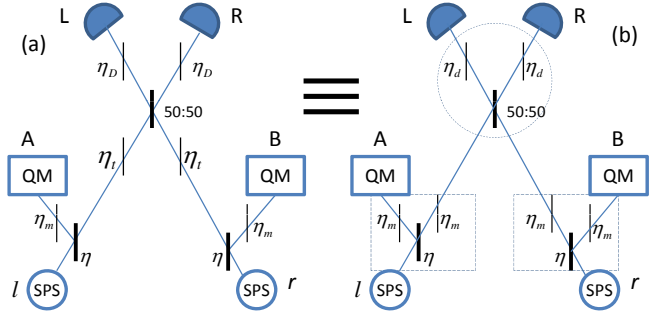}
\caption{{\footnotesize \label{fig:Model of setup} (Color online) A schematic model for the
SPS scheme. In (a) the memories' writing efficiencies, the path loss
and the detectors' efficiencies are represented by fictitious beam
splitters with transmission coefficients $\eta_{m},$ $\eta_{t}$
and $\eta_{D}$, respectively. In (b), an equivalent model is represented,
where we have grouped beam splitters in the form of butterfly modules;
see Fig. \ref{fig:Butterfly-transformation} . Here, $\eta_{t}\eta_{D}=\eta_{m}\eta_{d}$
and the model is valid so long as ${\eta_{t}\eta_{d}}\leqslant{\eta_{m}}.$}}
\end{figure}

In this section, we describe how we obtain parameters $P_{S},$ $P_{{\rm click}},$
and $R_{{\rm QKD}}$ for the setup in Fig. \ref{fig:Model of setup}(a)
and QKD measurements as in Fig. \ref{fig:QKD-measurements}.

Figure \ref{fig:Model of setup}(a) depicts the entanglement-distribution
setup for the SPS scheme. In our model the memories' writing efficiencies,
the path loss and the detectors' efficiencies are represented by fictitious
beam splitters with transmission coefficients $\eta_{m},$ $\eta_{t}$
and $\eta_{D},$ respectively, where $\eta_{t}=\exp[-L/(2L_{{\rm att}})]$
with $L_{{\rm att}}=25$~km for an optical fiber channel. Photodetectors,
in Fig. \ref{fig:Model of setup}, are then assumed to have unity
quantum efficiencies.

\begin{figure}
\begin{centering}
\includegraphics[width=3cm]{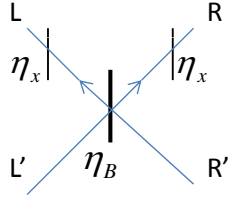} 
\par\end{centering}
\caption{{\footnotesize \label{fig:Butterfly-transformation} (Color online) A generic butterfly
module, represented by $B_{\eta_{B},\eta_{x}},$ where $\eta_{B}$
and $\eta_{x}$ are transmissivities for beam splitters shown in the
figure.}}
\end{figure}

In our analysis, we use an equivalent setup, as shown in Fig. \ref{fig:Model of setup}(b),
where beam splitters have been rearranged such that $\eta_{t}\eta_{D}=\eta_{m}\eta_{d}.$
We can then recognize similar building blocks, which we referred to
by butterfly modules, in Fig. \ref{fig:Model of setup}(b). A butterfly
module, as shown in Fig. \ref{fig:Butterfly-transformation}, is a
two-input two-output building block consisting of three beam splitters.
For an input state $\rho_{L'R'}$ in Fig. \ref{fig:Butterfly-transformation},
we denote the output state on ports $L$ and $R$ by $B_{\eta_{B},\eta_{x}}\left(\rho_{L'R'}\right).$

We use well-known models for beam splitters \cite{A:1} to find output
density matrices for input states to a generic butterfly module. In
Appendix A, we find the relevant input-output relationships for the
states of interest. We use Maple 15 to simplify some of our analytical
results. We can then find, $\rho_{ALBR},$ the joint state of the
memories and the optical modes entering detectors $L$ and $R$ in
Fig. \ref{fig:Model of setup}(b) by applying the butterfly operation
three times, as follows 
\begin{equation}
\rho_{ALBR}=B_{0.5,\eta_{d}}\left(B_{\eta,\eta_{m}}\left(\rho_{l}^{\left(in\right)}\right)\otimes B_{\eta,\eta_{m}}\left(\rho_{r}^{\left(in\right)}\right)\right).\label{Eq:ro_ALBR}
\end{equation}

According to the SPS protocol, a click on exactly one of the detectors
$L$ or $R,$ in Fig.~\ref{fig:Model of setup}(b), would herald
the success of entanglement distribution. This process can be modeled
by applying proper measurement operators considering whether PNRDs
or NRPDs are used. For example, for a click on detector $L$, the
explicit form of the measurement operator is given by 
\begin{equation}
M=\begin{cases}
(1-d_{c})[|1\left\rangle _{LL}\left\langle 1|\otimes|0\left\rangle _{RR}\left\langle 0|\right.\right.\right.\right.+\\
\ \ \ \ \ \ \ \ d_{c}|0\left\rangle _{LL}\left\langle 0|\otimes|0\left\rangle _{RR}\left\langle 0|],\qquad\qquad PNRD\right.\right.\right.\right.\\
(1-d_{c})[(I_{L}-|0\left\rangle _{LL}\left\langle 0|)\right.\right.\otimes|0\left\rangle _{RR}\left\langle 0|\right.\right.+\\
\ \ \ \ \ \ \ \ d_{c}|0\left\rangle _{LL}\left\langle 0|\otimes|0\left\rangle _{RR}\left\langle 0|],\qquad\qquad NRPD\right.\right.\right.\right.
\end{cases}\label{eq: PNRD-NRPD}
\end{equation}
 where $I_{L}$ denotes the identity operator for the mode entering
the left detector \cite{PhysRevA.73.042303}, and $d_{c}$ is the
dark-count rate per gate width per detector.

After the measurement, the resulting joint state, $\rho_{AB},$ of
quantum memories is given by: 
\begin{equation}
\rho_{AB}=\frac{{\rm tr}_{L,R}\left(\rho_{ALBR}M\right)}{P},\label{eq:rho in}
\end{equation}
 where 
\begin{equation}
P={\rm tr}\left(\rho_{ALBR}M\right)=\frac{P_{S}\left(L\right)}{2}\label{eq:equazione 6}
\end{equation}
 is the probability that the conditioning event $M$ occurs. The last
equality is due to the symmetry assumption.

For QKD measurements, we assume two pairs of memories, $A$-$B$ and
$C$-$D$, are given in an initial state similar to that of Eq.~\eqref{eq:rho in}.
We use the scheme described in Fig.~\ref{fig:QKD-measurements} to
perform QKD measurements. For simplicity, we assume both users use
zero phase shifts; other cases can be similarly worked out in our
symmetric setup. In Fig.~\ref{fig:QKD-measurements}, the retrieval
efficiency and the quantum detectors efficiency are represented by
fictitious beam splitters with, respectively, transmission coefficient
$\eta_{c}$ and $\eta_{D}.$ It is again possible to remodel the setup
in Fig.~\ref{fig:QKD-measurements} as shown in Fig.~\ref{fig:BSM}(b),
and use the butterfly operation $B_{0.5,\eta_{s}}$, where $\eta_{s}=\eta_{c}\eta_{D}.$
The density matrix right before photodetection in Fig.~\ref{fig:QKD-measurements}
is then given by $B_{0.5,\eta_{s}}\left(B_{0.5,\eta_{s}}\left(\rho_{AB}\otimes\rho_{CD}\right)\right),$
where one of the $B$-operators is applied to modes $A$ and $C$,
and the other one to modes $B$ and $D.$ Using this state, we find
$P_{{\rm click}}$ and $\epsilon_{Q}$ as outlined in Appendix B.

Using Eq.~\eqref{eq:equazione 2}, the secure key generation rate
per memory, $R_{{\rm QKD}}$, in the no-repeater setup, is then lower
bounded by \cite{PhysRevA.82.032304}: 
\begin{equation}
R_{1}=\max\left[(1-2\, H(\epsilon_{Q}))\,\dfrac{P_{S}\left(L\right)}{2L/c}\, P_{{\rm click}}/2,\,0\right]\label{eq:R1}
\end{equation}
 where $\dfrac{P_{S}\left(L\right)}{2L/c}$, given by Eq.~\eqref{eq:equazione 6},
is the generation rate of entangled pairs per logical memory $P_{{\rm click}}$
is the probability of creating a sifted key bit by using two entangled
pairs, and $[1-2\, H(\epsilon_{Q})]$ is the probability of creating
a secure key bit out of each sifted key bit. Here, we assume a biased basis choice to avoid an extra factor of two reduction in the rate \cite{Lo:EffBB84:2005}. The full definition for
$P_{{\rm click}}$ is given by Eq.~\eqref{eq:equazione B4}. The
QBER, 
\begin{equation}
\epsilon_{Q}=\cfrac{P_{{\rm error}}}{P_{{\rm click}}},
\end{equation}
 where $P_{{\rm error}}$ is the probability that Alice and Bob assign
different bits to their sifted keys, is given by Eq. \eqref{eq:equazione B5}.

\subsection{Repeater case}

First, consider the repeater setup of nesting level one in Fig.~\ref{fig:BSM}(a). We use the structure of Fig.~\ref{fig:Model of setup}(a)
to distribute entanglement between $A$-$A'$ and $B'$-$B$ memories.
The initial joint state of the system, $\rho_{AA'BB'}=\rho_{AA'}\otimes\rho_{BB'},$
can then be found, using Eq. \eqref{eq:rho in}, as described in the
previous section. We then apply a BSM by reading memories $A'$ and
$B'$ and interfering the resulting optical modes at a 50:50 beam
splitter. Success is declared if exactly one of the detectors in Fig.~\ref{fig:BSM}(a)
clicks. This can be modeled by applying measurement operators in Eq.~\eqref{eq: PNRD-NRPD},
which results in 
\begin{equation}
\rho_{AB}=\frac{{\rm tr}_{LR}\left(M\rho'_{ALBR}\right)}{P_{L}},\label{Eq:repstate}
\end{equation}
 where $\rho'_{ALBR}=B_{0.5,\eta_{s}}\left(\rho_{AA'BB'}\right)$,
where $L$ and $R$ represent the input modes to the photodetectors.
Note that, in Fig.~\ref{fig:BSM}, the detectors have ideal unity
quantum efficiencies. Moreover, 
\begin{equation}
P_{L}={\rm tr}\left(M\rho'_{ALBR}\right)=P_{M}/2
\end{equation}
 is the probability that only the left detector clicks in the BSM
module of Fig.~\ref{fig:BSM}. A click on the right detector has
the same probability by symmetry.

In order to find the secret key generation rate, we will follow similar
steps to the no-repeater case. That is, we apply the butterfly operation
to find relevant density matrices, from which $P_{{\rm click}}$ and
$\epsilon_{Q}$ can be obtained. From Eq.~\eqref{eq:equazione 2},
in the one-node repeater case, $R_{{\rm QKD}}$ is lower bounded by
\begin{equation}
R_{2}=\max\left[(1-2\, H(\epsilon_{Q}))\,\dfrac{P_{S}\left(L/2\right)}{2L/c}\, P_{M}\, P_{{\rm click}}/2,\,0\right].\label{eq:R2}
\end{equation}

Using the same approach, and by using Eq.~\eqref{eq:equazione 2}, we find the secret key generation rate for higher nesting levels. The details of which, have, however, been omitted for the sake of brevity.

\section{Numerical results}

In this section, we present numerical results for the secret key generation
rate of the SPS protocol, versus different system parameters, in the
no-repeater and repeater cases, and we compare them with
that of the DLCZ protocol. As mentioned earlier, we have used Maple
15 to analytically derive expressions for Eqs.~\eqref{eq:equazione 2}, \eqref{eq:R1}, and
\eqref{eq:R2} when PNRDs or NRPDs are used. Unless otherwise noted,
we use the nominal values summarized in Table \ref{tab:Values-of-inefficiencies}
for all the results presented in this section. 
\begin{table}
\begin{centering}
\begin{tabular}{|c|c|}
\hline 
Memory writing efficiency, $\eta_{m}$  & 0.5\tabularnewline
\hline 
Quantum efficiency, $\eta_{D}$  & 0.3\tabularnewline
\hline 
Memory retrieval efficiency, $\eta_{c}$  & 0.7\tabularnewline
\hline 
Dark count per pulse, $d_{c}$  & $10^{-6}$\tabularnewline
\hline 
Attenuation length, $L_{att}$  & 25 km\tabularnewline
\hline 
Speed of light, $c$  & $2\cdot10^{5}$ km/s\tabularnewline
\hline 
Decay (dephasing) time constants, $T_{1}\,(T_{2})$  & $\infty$ \tabularnewline
\hline 
\end{tabular}
\par\end{centering}

\caption{{\footnotesize \label{tab:Values-of-inefficiencies}Nominal values
used in our numerical results.}}
\end{table}

\subsection{SPS key rate versus system parameters}

\subsubsection{Source transmission coefficient}

\begin{figure}
\begin{centering}
\includegraphics[width=8.6cm]{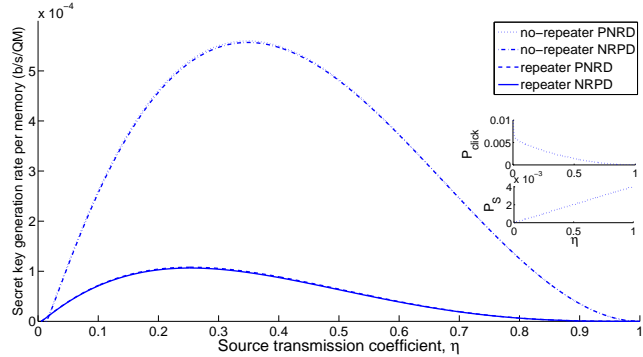}
\par\end{centering}
\caption{{\footnotesize \label{fig:Rqkd_vs_eta_norep} (Color online) $R_{\rm QKD}$ versus the
source transmission coefficient $\eta$ for the PNRDs and NRPDs in
the no-repeater and one-node repeater cases. Here, $p=0.001$, $L=250$~km, and $n=1$ for the repeater system; other parameters are listed in Table~\ref{tab:Values-of-inefficiencies}}}
\end{figure}

Figure \ref{fig:Rqkd_vs_eta_norep} shows the secret key generation
rate per memory, $R_{{\rm QKD}}$, versus the source transmission
coefficient $\eta$ in Fig.~\ref{fig:DLCZ and SPS}(b), at $p=0.001$
and $L=250$ km. It can be seen that there exist optimal values of
$\eta$ for both repeater and no-repeater systems. Table~\ref{Tab:Opt}
summarizes these optimum values for different nesting levels. The optimal value
of $\eta$ for the no-repeater system is higher than the repeater
ones, and that is because of the additional entanglement swapping steps
in the latter systems. Another remarkable feature in Fig.~\ref{fig:Rqkd_vs_eta_norep}
is that the penalty of using NRPDs, versus PNRDs, seems to be minor
at $p=10^{-3}$. PNRDs better show their advantage at higher values
of $p$ when double-photon terms become more evident.

The existence of an optimal value for $\eta$ arises from a competition
between the probability of entanglement distribution $P_{S}$, which
grows with $\eta,$ and $P_{{\rm click}},$ which decreases with $\eta$.
This has been demonstrated in the inset of Fig.~\ref{fig:Rqkd_vs_eta_norep}.
The latter issue is mainly because of the vacuum component in Eq.~\eqref{Eq:SPSout}.
In the case of the repeater system, $P_{M}$ also decreases with $\eta$
for the same reason, and that is why the optimal value of $\eta$
is lower for repeater systems.

\begin{table}
\begin{centering}
\begin{tabular}{c|c|c}
 nesting level & PNRD  & NRPD \tabularnewline
\hline 
 0  & 0.35  & 0.34 \tabularnewline
\hline 
 1  & 0.28  & 0.27 \tabularnewline
\hline 
 2  & 0.21  & 0.20* \tabularnewline
\hline 
 3  & 0.12  & 0.11* \tabularnewline
\end{tabular}
\par\end{centering}
\caption{{\footnotesize \label{Tab:Opt} Optimal values of $\eta$, at $p=0.001$
and $L=250$ km, for repeater and no-repeater systems, when PNRDs
or NRPDs are used. The figures with an asterisk are approximate values.}}
\end{table}

The optimum values of $\eta$ in Fig.~\ref{fig:Rqkd_vs_eta_norep}
are interestingly almost identical to the value of $\eta$ that minimizes
the total time for a successful creation of an entangled state, as
prescribed in \cite{PhysRevA.76.050301}. It is because, at a fixed
distance, the QBER term in Eqs.~\eqref{eq:R1} and \eqref{eq:R2}
is mainly a function of the double-photon probability and the dark
count rate, and it does not considerably vary with $\eta$. More generally, the optimum values of $\eta$ remain constant as in
Table~\ref{Tab:Opt} so long as the error terms are well below the
cut-off threshold in QKD.


\subsubsection{Nesting levels and crossover distance}

\begin{figure}
\begin{centering}
\includegraphics[width=8.6cm]{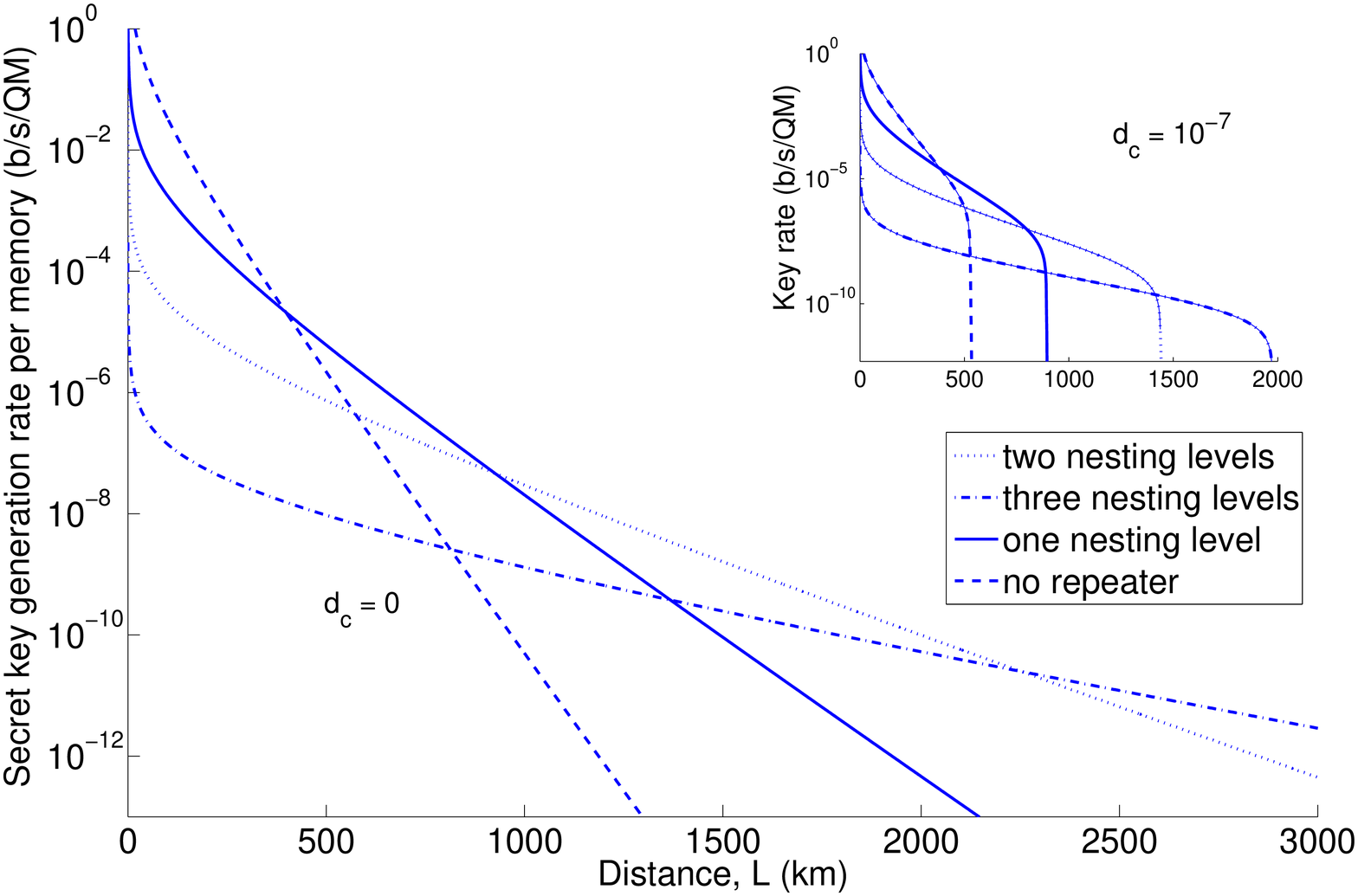}
\par\end{centering}
\caption{{\footnotesize \label{fig:R vs eta diff L} (Color online)  $R_{\rm QKD}$ versus distance for up to three nesting levels at two different dark count rates at $p=10^{-4}$. All other values are listed in Tables~\ref{tab:Values-of-inefficiencies} and \ref{Tab:Opt}.}}
\end{figure}

Figure \ref{fig:R vs eta diff L} depicts the normalized secret key generation rate versus distance for different nesting levels. At $d_c = 0$, the slope advantage, proportional to $P_S(L/2^n)$, for higher nesting levels is clear in the figure. Because of additional entanglement swapping stages, the no-path-loss rate at $L=0$ is, however, lower for higher nesting levels. That would result in crossover distances---at which one system outperforms another---once we move from one nesting level to its subsequent one. The crossover distance has architectural importance and will specify the optimum distance between repeater nodes.

The crossover distance is a function of various system parameters. As shown in the inset of Fig.~\ref{fig:R vs eta diff L}, positive dark count rates can considerably change the crossover distance. By including dark counts in our analysis, there will be a cutoff security distance for each nesting level. By increasing the dark count rate, these cutoff distances will decrease and become closer to each other. That would effectively reduce the crossover distance. At dark count rates as high as $d_c = 10^{-6}$, the superiority of 3 over 2 nesting levels at long distances would almost diminish as they both have almost the same cutoff distances.

\begin{figure}
\begin{centering}
\includegraphics[width=8.6cm]{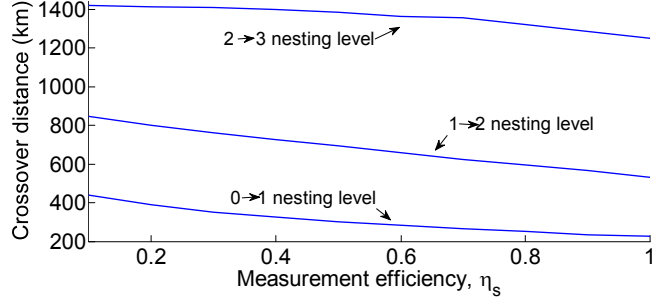}
\par\end{centering}
\caption{{\footnotesize \label{fig:Rate--dist} (Color online) The crossover distance, at which a repeater system with nesting level $n$ outperforms a system with nesting level $n-1$,
as a function of measurement efficiency $\eta_s = \eta_{c}\eta_{D}$, at $p=10^{-4}$. All other parameters are taken from Tables~\ref{tab:Values-of-inefficiencies} and \ref{Tab:Opt} except for the dark count, which is $10^{-7}$. }}
\end{figure}

The crossover distance will decrease if component
efficiencies go up. This has been shown in Fig.~\ref{fig:Rate--dist} when the crossover distance is depicted versus measurement
efficiency. The latter directly impacts the BSM success probability,
$P_{M}$, and that is why the larger its value the lower the
crossover distance. Larger values of $\eta_{m}$ also reduce the vacuum
component, thus enhancing the chance of success at the entanglement
swapping stage.

It can be noted in Fig.~\ref{fig:Rate--dist} that, even for highly efficient devices, 
the optimum distance between repeater nodes would tend to lie at around 150-200~km. For instance at $L=1000$~km, and with the nominal values used in this paper, the optimum nesting level is 2, which
implies that the distance between two nodes of the repeater is
250~km. This could be a long distance for practical
purposes, such as for phase stabilization, and that might require
us to work at a suboptimal distancing. The latter would further reduce
the secret key generation rate. Our result is somehow different from what is reported in \cite{RevModPhys.83.33,Razavi_SPIE}, albeit one should bear in mind the different set of assumptions and measures used therein. 

\subsubsection{Double-photon probability}

\begin{figure}
\begin{centering}
\includegraphics[width=8.6cm]{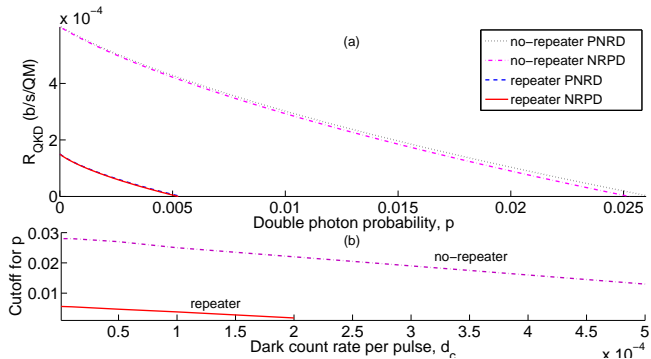}
\par\end{centering}
\caption{{\footnotesize \label{fig:Rqkd_vs_p_norep}(Color online) (a) $R_{\rm QKD}$ versus double-photon
probability, $p$, using PNRDs and NRPDs in the no-repeater and one-node repeater
cases. (b) Cutoff double-photon probability, at which the key rate becomes zero, versus the dark count rate $d_{c}$. The higher the dark count rate, the less room for multi-photon errors. All graphs are at $L=250$ km.}}
\end{figure}

Figures \ref{fig:Rqkd_vs_p_norep} show the secret key generation
rate for the SPS protocol, at the optimal values of $\eta$ listed in Table~\ref{Tab:Opt}, versus
the double-photon probability $p$ in the no-repeater and repeater
cases. It can be seen that, in both cases, there exists a cutoff probability
at which $R_{{\rm QKD}}$ becomes zero. This point corresponds to
the threshold QBER of $11\%$ from the Shor-Preskill security proof.
In the case of QMs with sufficiently long coherence times, as is the
case in Fig.~\ref{fig:Rqkd_vs_p_norep}, the QBER in our system stems
from two factors: dark count and double-photon probability. The former
is proportional to $d_{c}/\eta_{d}$ and it comes into effect only
when the path loss is significant. The latter, however, affects the
QBER at all distances. To better see this issue, in Fig.~\ref{fig:Rqkd_vs_p_norep}(b),
the cutoff probability is depicted versus the dark count rate. It
can be seen that the cutoff probability linearly goes down with $d_{c}$,
which confirms the additive contribution of dark counts and two-photon
emissions to the QBER.

The cutoff probability at $d_{c}=0$ deserves a particular attention.
As can be seen in Fig.~\ref{fig:Rqkd_vs_p_norep}(b), for the no-repeater
system, the maximum allowed value of $p$ is about $0.028$ for PNRDs
and $0.026$ for NRPDs. This implies that the QBER in this case, at
$d_{c}=0$, is roughly given by $4p$. This can be verified by finding the contributions from 
two- and single-photon components in Eq.~\eqref{eq:input_state}. We can then show that the QBER, at the optimal value of
$\eta$ in Table~\ref{Tab:Opt}, is roughly given by $3(1+\eta)p \approx 4p$. Similarly, in the repeater case, one can
show that each BSM almost doubles the contribution of two-photon emissions
to the QBER. Considering that four pairs of entangled states is now needed, and that the chance of making an error for an unentangled
pair is typically 1/2, the QBER is roughly given by $4\times2\times3(1+\eta)p/2\approx16p$,
which implies that, to the first-order approximation, the maximum
allowed value for $p$ is about $0.11/16=0.0068$. Figure~\ref{fig:Rqkd_vs_p_norep}(a)
confirms this result, where the cutoff probability is about $0.0056$
for the PNRDs and $0.0054$ for the NRPDs, corresponding to $\epsilon_Q \approx 20p$.

With a similar argument as above, one may roughly expect a factor of 4-to-5
increase in the QBER for each additional nesting level. This implies
that for a repeater system with nesting level 3, we should expect
a QBER around $500p$ just because of the double-photon emission. Table~\ref{tab:cutoff_prob} confirms our approximation by providing the actual cutoff figures for different nesting levels. We discuss the practical implications of this finding later in this
section.

\begin{table}
\begin{centering}
\begin{tabular}{|c|c|}
\hline 
nesting level & cutoff double-photon probability\tabularnewline
\hline 
0  & $2.5 \times 10^{-2}$\tabularnewline
\hline 
1  & $5.0 \times 10^{-3}$\tabularnewline
\hline 
2  & $1.8 \times 10^{-3}$\tabularnewline
\hline 
3  & $2.1 \times 10^{-4}$\tabularnewline
\hline 
\end{tabular}
\par\end{centering}

\caption{{\footnotesize \label{tab:cutoff_prob}Cutoff double-photon probabilities when PNRDs are used for different nesting levels. The paramter values used are listed in Tables~\ref{tab:Values-of-inefficiencies} and \ref{Tab:Opt} .}}
\end{table}

\subsubsection{Memory dephasing}

\begin{figure}
\begin{centering}
\includegraphics[width=8.6cm]{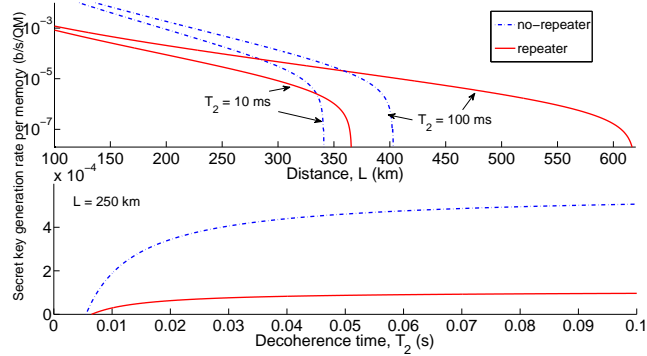}
\par\end{centering}
\caption{\label{fig:Decoherence-time}{\footnotesize (Color online) (a) The secret key
generation rate versus distance for two values of decoherence time, $T_{2}=10$~ms and $100$~ms. In (b) the
secret key rate is plotted as a function of $T_{2}$ at $L=250$~km. In both graphs, $p=10^{-3}$. }}
\end{figure}

Figure \ref{fig:Decoherence-time}(a) shows the secret key generation
rate per memory for the SPS protocol with NRPDs versus distance for two different values
of the dephasing time, $T_{2},$ at $p=10^{-3}.$ It is clear that, by reducing the coherence time, the security distance drops to shorter distances. Whereas, at $T_2 = 100$~ms, the key rate remains the same as that of Fig.~\ref{fig:R vs eta diff L}(b), at $T_2 = 10$~ms, both repeater and non-repeater systems would fall short of supporting distances over 360~km.

Figure \ref{fig:Decoherence-time}(b) shows the secret key generation rate per memory versus $T_2$ at $L=250$~km. There is a minimum required coherence time of around 5~ms below which we cannot exchange a secure key. This point corresponds to the 11\% QBER mainly caused by the dephasing process. In fact, at this point, we have $\epsilon_Q \approx e_d = (1-\exp[-L/(cT_2)])/2 = 0.11$, which implies that the maximum distance supported by our protocol is about $c T_2/4$. To be operating on the flat region in the curves shown in Fig.~\ref{fig:Decoherence-time}(b), one even requires a higher coherence time. In other words, the minimum required coherence time to support a link of length $L$ is on the order of $10 L/c$. This is in line with findings in \cite{PhysRevA.80.032301}. Although not explicitly shown here, the same requirements are expected to be as well applicable to other QKD systems that rely on quantum repeaters.

\subsection{SPS versus DLCZ}

\begin{figure}
\begin{centering}
\includegraphics[width=8.6cm]{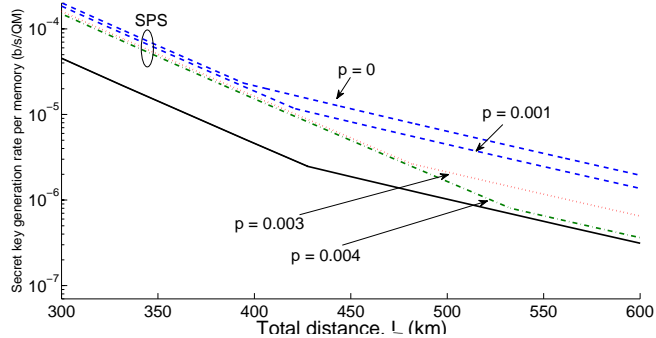}
\par\end{centering}
\caption{{\footnotesize \label{fig:Comparison-between-DLCZ} (Color online) Comparison between
the DLCZ and SPS protocols using PNRDs. For both systems, the better of repeater or non-repeater system is used. Both systems operate at their optimal setting: For the SPS protocol, the optimum value of $\eta$ is used; for the DLCZ protocol, the optimum value of $p_c$ is used. By varying the double-photon probability, $p$, in the SPS protocol, we find that the maximum $p$ at which SPS outperforms DLCZ is around $p=0.004$. In all curves, $d_c=0$. All other parameters are taken from Tables~\ref{tab:Values-of-inefficiencies} and \ref{Tab:Opt}.}}
\end{figure}

Figure \ref{fig:Comparison-between-DLCZ} compares the secret key generation rate for the SPS protocol, found in this paper, with that of the DLCZ protocol as obtained in \cite{PhysRevA.82.032304}. In both systems, we have assumed $d_{c}=0$. All other parameters are
as in Table~\ref{tab:Values-of-inefficiencies}. In both systems, we use the optimal setting in the PNRD case. The conclusion would be similar if one uses NRPDs, as seen in all numerical results presented in this paper. For the SPS protocol, the optimal setting corresponds to the values of $\eta$ in Table~\ref{Tab:Opt}. In the DLCZ protocol, the adjustable parameter is the excitation probability $p_c$. Note that, whereas in the SPS protocol, the rate decreases monotonically with $p$, in the DLCZ protocol, it peaks at a certain value of $p_c$. That is because, in the SPS protocol, we use an on-demand source of photons, whereas in the DLCZ protocol, the heralding probability as well as the relative double-photon probability are both proportional to $p_c$. The optimum value for the excitation probability is given by $p_{c}=0.0243$ in the no-repeater case and $p_{c}=0.0060$ in the one-node repeater case \cite{PhysRevA.82.032304}. Note that the analysis in \cite{PhysRevA.82.032304} accounts for all multi-excitation components in the initial state of the system.  In all curves in Fig.~\ref{fig:Comparison-between-DLCZ}, we have used the better of the repeater and no repeater systems at each distance. Our results show that the SPS protocol offers a higher key rate per memory than the
DLCZ for on-demand single-photon sources with double-photon probabilities of 0.004 or lower. The advantage is however below one order of magnitude
in most cases.

A key assumption in the results obtained above is the use of on-demand sources in the SPS protocol. The less-than one-order-of-magnitude
difference between the two protocols can then be easily washed away if one uses single-photon sources with less than roughly 50\% efficiencies. This means that the conventional methods for generating single photons, such as parametric downconversion or quantum dots, may not yet be useful in the SPS protocol. The partial memory-readout technique could, still, be a viable solution. In this scheme, we drive a Raman transition, as in the DLCZ protocol, in an atomic ensemble, such that with some probability $p$ a Stokes photon is released. If we detect such a
photon, then we are left with an ensemble, which can be partially read out with probability $\eta$ to resemble the first part of the SPS protocol. One should, however, note that with limitations on the cutoff probability to be on the order of $10^{-4}$--$10^{-5}$, it may take quite a long time to prepare such a source-memory pair. For instance, if the required $p$ is $10^{-4}$, and the efficiency of the collection and detection setup is 0.1, even if we run the driving pulse at a 1~GHz rate, it takes on average 0.1~ms to prepare the initial state. This time is comparable to the time that it takes for light to travel 100~km, which is on the same order of magnitude that we run our cyclic protocol in Fig.~\ref{fig: mult mem conf}(a). Considering a particular setup paramters, it is not then an obvious call to which of the DLCZ or SPS protocols performs better, and that underlines
the importance of our theoretical analysis.

\section{Conclusions}

In this paper, we analyzed the SPS protocol proposed in \cite{PhysRevA.76.050301} in terms of the secret key generation rate that it could offer in a QKD-over-repeater setup. This protocol belongs to a family of probabilistic quantum repeaters, and perhaps one of their best, inspired by the DLCZ proposal \cite{:1}. Our aim was to compare the SPS protocol, for QKD applications, with the original DLCZ protocol, as reported in \cite{PhysRevA.82.032304}, in a realistic scenario. To this end, we considered various sources of imperfections in our analysis and obtained the optimal regime of operation as a function of system parameters. We accounted for double-photon probabilities at the source and realized that, under Shor-Preskill's security-proof assumptions, its value should not exceed 0.11/4, in a direct-link scenario, and 0.11/20 in a one-node repeater case. We would expect the same scaling, if not worse, at higher nesting levels, which implied that for a repeater setup of nesting level 3, the double-photon probability must be on the order of $10^{-4}$ or lower. That would be a challenging requirement for on-demand single-photon sources needed in the SPS protocol. Under above circumstances, the advantage of the SPS protocol over the DLCZ would be marginal and would not exceed one order of magnitude of key rate in bit/s per memory. In our analysis, we also accounted for memory dephasing and dark counts. Our results showed that the minimum required coherence time for a link of length $L$ is roughly given by $4L/c$, where $c$ is the speed of light in the channel. The crossover distance at which we have to move up the nesting-level ladder varies for different system parameters. The optimum distancing between repeater nodes can nevertheless be typically as high as  150~km to 200~km depending on the measurement efficiency among other parameters. We noticed that, within practical regimes of operation, there would only be a minor advantage in using resolving photodetectors over more conventional threshold detectors. We emphasized that, because of using a normalized figure of merit in our analysis, our results would be applicable to multi-memory and/or -mode scenarios.

\section*{Acknowledgments}
The authors would like to thank X. Ma for fruitful discussions. This work was in part supported by the European Community's Seventh Framework Programme under Grant Agreement 277110 and the UK Engineering and Physical Science Research Council grant number EP/J005762/1.

\appendix
\section{Butterfly transformation}

\label{Sec:Buterfly} In this Appendix, we find input-output relationships
for the butterfly module in Fig.~\ref{fig:Butterfly-transformation}.
We do this in the number-state representation, only for the relevant
input states in Eq.~\eqref{Eq:ro_ALBR}.

\begin{table}
\begin{footnotesize} %
\begin{tabular}{|c|c|}
\hline 
$\rho_{in}$  & $B_{\eta,\eta_{m}}\left(\rho_{in}\right)$\tabularnewline
\hline 
\hline 
$|10\left\rangle \left\langle 10|\right.\right.$  & $\eta\eta_{m}|01\left\rangle \left\langle 01|\right.\right.+\eta_{m}\sqrt{\eta\left(1-\eta\right)}\left(|10\left\rangle \left\langle 01|\right.\right.+|01\left\rangle \left\langle 10|\right.\right.\right)+\eta_{m}\left(1-\eta\right)|10\left\rangle \left\langle 10|\right.\right.+\left(1-\eta_{m}\right)|00\left\rangle \left\langle 00|\right.\right.$\tabularnewline
\hline 
 & $(1-\eta_{m})^{2}|00\left\rangle \left\langle 00|\right.\right.+2\eta\eta_{m}(1-\eta_{m})|01\left\rangle \left\langle 01|\right.\right.+\eta\eta_{m}^{2}(1-\eta)(|20\left\rangle \left\langle 02|\right.+\right.|02\left\rangle \left\langle 20|)\right.\right.$\tabularnewline
$|20\left\rangle \left\langle 20|\right.\right.$  & $+2\eta_{m}(1-\eta_{m})\sqrt{\eta(1-\eta)}(|10\left\rangle \left\langle 01|\right.+|01\left\rangle \left\langle 10|\right.\right.)+\eta^{2}\eta_{m}^{2}|02\left\rangle \left\langle 02|\right.\right.\right.+2\eta\eta_{m}^{2}(1-\eta)|11\left\rangle \left\langle 11|\right.\right.$\tabularnewline
 & $+\eta_{m}^{2}(1-\eta)\sqrt{2\eta(1-\eta)}(|20\left\rangle \left\langle 11|\right.\right.+|11\left\rangle \left\langle 20|\right.\right.)+\eta\eta_{m}^{2}\sqrt{2\eta(1-\eta)}(|02\left\rangle \left\langle 11|\right.\right.+|11\left\rangle \left\langle 02|\right.\right.)$ \tabularnewline
 & $+2\eta_{m}(1-\eta)(1-\eta_{m})|10\left\rangle \left\langle 01|\right.\right.+\eta_{m}^{2}(1-\eta)^{2}|20\left\rangle \left\langle 20|\right.\right.$\tabularnewline
\hline 
\end{tabular}\caption{The input-output relationship for the $B_{\eta,\eta_{m}}$ operator. $|jk\rangle \langle jk| = |j\rangle_{JJ} \langle j| \otimes |k\rangle_{KK} \langle k|$, where $J=L'$ and $K=R'$ for input number states and $J=L$ and $K=R$ for output number states in Fig.~\ref{fig:Butterfly-transformation}.} 
\label{Tab:Beta_etam} \end{footnotesize} 
\end{table}

Table \ref{Tab:Beta_etam} provides the output state for the butterfly
operation $B_{\eta,\eta_{m}}$ when there is exactly one or two photons at one of the input ports. These are the only relevant terms in the
input states in Eqs.~\eqref{eq:initial density matrix} and \eqref{eq:input_state}.
Using Table \ref{Tab:Beta_etam}, we find $B_{\eta,\eta_{m}}(\rho_{l}^{(in)})\otimes B_{\eta,\eta_{m}}(\rho_{r}^{(in)})$,
to be used in Eq.~\eqref{Eq:ro_ALBR}.

The last operation required in Eq.~\eqref{Eq:ro_ALBR} is the symmetric
butterfly operation $B_{0.5,\eta_{d}}$. Table~\ref{Tab:B05eta}
lists the input-output relationships for all relevant input terms
in our system for the more general operation $B_{0.5,\eta_{x}}$.
Note that by choosing $\eta_{x}=\eta_{s}$, we can use the same relationships
for the measurement modules used in entanglement swapping and QKD
of Figs.~\ref{fig:BSM} and \ref{fig:QKD-measurements}, respectively.
For the sake of brevity, in Table~\ref{Tab:B05eta}, we have only
included the terms that provide us with nonzero values after applying
the measurement operation. More specifically, we have removed all
{\em asymmetric} density matrix terms, such as $|10\left\rangle \left\langle 01|\right.\right.$
or $|01\left\rangle \left\langle 10|\right.\right.$, for which the
bra state is different from the ket state, from the output state.

\begin{table}
\begin{footnotesize} %
\begin{tabular}{|c|c|}
\hline 
$\rho_{in}$  & $B_{0.5,\eta_{x}}\left(\rho_{in}\right)$\tabularnewline
\hline 
\hline 
$|10\left\rangle \left\langle 10|\right.\right.$  & $\frac{\eta_{x}}{2}\left(|10\left\rangle \left\langle 10|\right.\right.+|01\left\rangle \left\langle 01|\right.\right.\right)+\left(1-\eta_{x}\right)|00\left\rangle \right\langle 00|$\tabularnewline
\hline 
$|01\left\rangle \left\langle 01|\right.\right.$  & $\frac{\eta_{x}}{2}\left(|10\left\rangle \left\langle 10|\right.\right.+|01\left\rangle \left\langle 01|\right.\right.\right)+\left(1-\eta_{x}\right)|00\left\rangle \right\langle 00|$\tabularnewline
\hline 
$|11\left\rangle \left\langle 11|\right.\right.$  & $\eta_{x}\left(1-\eta_{x}\right)\left(|10\left\rangle \left\langle 10|\right.\right.+|01\left\rangle \left\langle 01|\right.\right.\right)+\left(1-\eta_{x}\right)^{2}|00\left\rangle \right\langle 00|+\frac{\eta_{x}^{2}}{2}\left(|20\left\rangle \left\langle 20|\right.\right.+|02\left\rangle \left\langle 02|\right.\right.\right)$\tabularnewline
\hline 
$|20\left\rangle \left\langle 20|\right.\right.$  & $\eta_{x}\left(1-\eta_{x}\right)\left(|10\left\rangle \left\langle 10|\right.\right.+|01\left\rangle \left\langle 01|\right.\right.\right)+\left(1-\eta_{x}\right)^{2}|00\left\rangle \right\langle 00|+\frac{\eta_{x}^{2}}{2}|11\left\rangle \left\langle 11|\right.\right.+\frac{\eta_{x}^{2}}{4}\left(|20\left\rangle \left\langle 20|\right.\right.+|02\left\rangle \left\langle 02|\right.\right.\right)$\tabularnewline
\hline 
$|02\left\rangle \left\langle 02|\right.\right.$  & $\eta_{x}\left(1-\eta_{x}\right)\left(|10\left\rangle \left\langle 10|\right.\right.+|01\left\rangle \left\langle 01|\right.\right.\right)+\left(1-\eta_{x}\right)^{2}|00\left\rangle \right\langle 00|+\frac{\eta^{2}}{2}|11\left\rangle \left\langle 11|\right.\right.+\frac{\eta_{x}^{2}}{4}\left(|20\left\rangle \left\langle 20|\right.\right.+|02\left\rangle \left\langle 02|\right.\right.\right)$\tabularnewline
\hline 
$|21\left\rangle \left\langle 21|\right.\right.$  & $\frac{3}{2}\eta_{x}\left(1-\eta_{x}\right)^{2}\left(|10\left\rangle \left\langle 10|\right.\right.+|01\left\rangle \left\langle 01|\right.\right.\right)+\left(1-\eta_{x}\right)^{3}|00\left\rangle \right\langle 00|+\frac{\eta_{x}^{2}}{2}\left(1-\eta_{x}\right)|11\left\rangle \left\langle 11|\right.\right.$\tabularnewline
 & $+\frac{5}{4}\eta_{x}^{2}\left(1-\eta_{x}\right)\left(|20\left\rangle \left\langle 20|\right.\right.+|02\left\rangle \left\langle 02|\right.\right.\right)+\frac{3}{8}\eta_{x}^{3}\left(|30\left\rangle \left\langle 30|\right.\right.+|03\left\rangle \left\langle 03|\right.\right.\right)+\frac{1}{8}\eta_{x}^{3}\left(|21\left\rangle \left\langle 21|\right.\right.+|12\left\rangle \left\langle 12|\right.\right.\right)$\tabularnewline
\hline 
$|21\left\rangle \left\langle 21|\right.\right.$  & $\frac{3}{2}\eta_{x}\left(1-\eta_{x}\right)^{2}\left(|10\left\rangle \left\langle 10|\right.\right.+|01\left\rangle \left\langle 01|\right.\right.\right)+\left(1-\eta_{x}\right)^{3}|00\left\rangle \right\langle 00|+\frac{\eta_{x}^{2}}{2}\left(1-\eta_{x}\right)|11\left\rangle \left\langle 11|\right.\right.$\tabularnewline
 & $+\frac{5}{4}\eta_{x}^{2}\left(1-\eta_{x}\right)\left(|20\left\rangle \left\langle 20|\right.\right.+|02\left\rangle \left\langle 02|\right.\right.\right)+\frac{3}{8}\eta_{x}^{3}\left(|30\left\rangle \left\langle 30|\right.\right.+|03\left\rangle \left\langle 03|\right.\right.\right)+\frac{1}{8}\eta_{x}^{3}\left(|21\left\rangle \left\langle 21|\right.\right.+|12\left\rangle \left\langle 12|\right.\right.\right)$\tabularnewline
\hline 
$|10\left\rangle \left\langle 01|\right.\right.$  & $\frac{1}{2}\eta_{x}\left(|10\left\rangle \left\langle 10|\right.\right.-|01\left\rangle \left\langle 01|\right.\right.\right)$\tabularnewline
\hline 
$|01\left\rangle \left\langle 10|\right.\right.$  & $\frac{1}{2}\eta_{x}\left(|10\left\rangle \left\langle 10|\right.\right.-|01\left\rangle \left\langle 01|\right.\right.\right)$\tabularnewline
\hline 
$|11\left\rangle \left\langle 20|\right.\right.$  & $\frac{\sqrt{2}}{2}\eta_{x}\left(1-\eta_{x}\right)\left(|10\left\rangle \left\langle 10|\right.\right.-|01\left\rangle \left\langle 01|\right.\right.\right)+\frac{1}{2\sqrt{2}}\eta_{x}^{2}\left(|20\left\rangle \left\langle 20|\right.\right.-|02\left\rangle \left\langle 02|\right.\right.\right)$\tabularnewline
\hline 
$|11\left\rangle \left\langle 02|\right.\right.$  & $\frac{\sqrt{2}}{2}\eta_{x}\left(1-\eta_{x}\right)\left(|10\left\rangle \left\langle 10|\right.\right.-|01\left\rangle \left\langle 01|\right.\right.\right)+\frac{1}{2\sqrt{2}}\eta_{x}^{2}\left(|20\left\rangle \left\langle 20|\right.\right.-|02\left\rangle \left\langle 02|\right.\right.\right)$\tabularnewline
\hline 
$|20\left\rangle \left\langle 11|\right.\right.$  & $\frac{\sqrt{2}}{2}\eta_{x}\left(1-\eta_{x}\right)\left(|10\left\rangle \left\langle 10|\right.\right.-|01\left\rangle \left\langle 01|\right.\right.\right)+\frac{1}{2\sqrt{2}}\eta_{x}^{2}\left(|20\left\rangle \left\langle 20|\right.\right.-|02\left\rangle \left\langle 02|\right.\right.\right)$\tabularnewline
\hline 
$|02\left\rangle \left\langle 11|\right.\right.$  & $\frac{\sqrt{2}}{2}\eta_{x}\left(1-\eta_{x}\right)\left(|10\left\rangle \left\langle 10|\right.\right.-|01\left\rangle \left\langle 01|\right.\right.\right)+\frac{1}{2\sqrt{2}}\eta_{x}^{2}\left(|20\left\rangle \left\langle 20|\right.\right.-|02\left\rangle \left\langle 02|\right.\right.\right)$\tabularnewline
\hline 
$|21\left\rangle \left\langle 21|\right.\right.$  & $\eta_{x}\left(1-\eta_{x}\right)^{2}\left(|10\left\rangle \left\langle 10|\right.\right.-|01\left\rangle \left\langle 01|\right.\right.\right)+\eta_{x}^{2}\left(1-\eta_{x}\right)\left(|20\left\rangle \left\langle 20|\right.\right.-|02\left\rangle \left\langle 02|\right.\right.\right)$\tabularnewline
 & $+\frac{3}{8}\eta_{x}^{3}\left(|30\left\rangle \left\langle 30|\right.\right.-|03\left\rangle \left\langle 03|\right.\right.\right)+\frac{1}{8}\eta_{x}^{3}\left(|12\left\rangle \left\langle 12|\right.\right.-|21\left\rangle \left\langle 21|\right.\right.\right)$\tabularnewline
\hline 
$|12\left\rangle \left\langle 12|\right.\right.$  & $\eta_{x}\left(1-\eta_{x}\right)^{2}\left(|10\left\rangle \left\langle 10|\right.\right.-|01\left\rangle \left\langle 01|\right.\right.\right)+\eta_{x}^{2}\left(1-\eta_{x}\right)\left(|20\left\rangle \left\langle 20|\right.\right.-|02\left\rangle \left\langle 02|\right.\right.\right)$\tabularnewline
 & $+\frac{3}{8}\eta_{x}^{3}\left(|30\left\rangle \left\langle 30|\right.\right.-|03\left\rangle \left\langle 03|\right.\right.\right)+\frac{1}{8}\eta_{x}^{3}\left(|12\left\rangle \left\langle 12|\right.\right.-|21\left\rangle \left\langle 21|\right.\right.\right)$\tabularnewline
\hline 
 & $\left(1-\eta_{x}\right)^{4}|00\left\rangle \right\langle 00|+2\eta_{x}\left(1-\eta_{x}\right)^{3}\left(|10\left\rangle \left\langle 10|\right.\right.+|01\left\rangle \left\langle 01|\right.\right.\right)+\eta_{x}^{2}\left(1-\eta_{x}\right)^{2}|11\left\rangle \right\langle 11|$\tabularnewline
$|22\left\rangle \left\langle 22|\right.\right.$  & $+\frac{3}{2}\eta_{x}^{3}\left(1-\eta_{x}\right)\left(|30\left\rangle \left\langle 30|\right.\right.+|03\left\rangle \left\langle 03|\right.\right.\right)+\frac{1}{2}\eta_{x}^{3}\left(1-\eta_{x}\right)\left(|21\left\rangle \left\langle 21|\right.\right.+|12\left\rangle \left\langle 12|\right.\right.\right)$\tabularnewline
 & $\frac{5}{2}\eta_{x}^{2}\left(1-\eta_{x}\right)^{2}\left(|20\left\rangle \left\langle 20|\right.\right.+|02\left\rangle \left\langle 02|\right.\right.\right)+\frac{3}{8}\eta_{x}^{4}\left(|40\left\rangle \left\langle 40|\right.\right.+|04\left\rangle \left\langle 04|\right.\right.\right)+\frac{1}{4}\eta_{x}^{4}|22\left\rangle \right\langle 22|$\tabularnewline
\hline 
\end{tabular}\label{Tab:B05eta} \caption{The input-output relationship for a symmetric butterfly module. The notation used is similar to that of Table~\ref{Tab:Beta_etam}.}
\end{footnotesize} 
\end{table}

\section{Derivation of $P_{{\rm click}}$ and $P_{{\rm error}}$}

In this Appendix, we find the gain and the QBER for the QKD scheme
of Fig.~\ref{fig:QKD-measurements}. Let us assume that the memory
pairs $AB$ and $CD$ are already entangled via the no-repeater or
the one-node repeater scheme described in Sec.~\ref{Sec:SPS}. In
the case of SPS protocol, their state is, respectively, given by Eqs.~\eqref{eq:rho in}
and \eqref{Eq:repstate}. The density matrix right before photodetection
in Fig.~\ref{fig:QKD-measurements} is then given by $\rho_{ABCD}=B_{0.5,\eta_{s}}\left(B_{0.5,\eta_{s}}\left(\rho_{AB}\otimes\rho_{CD}\right)\right),$
where one of the $B$-operators is applied to modes $A$ and $C$,
and the other one to modes $B$ and $D.$ Using Table~\ref{Tab:B05eta},
we can calculate the exact form of $\rho_{ABCD}$, as we have done
in this paper.

The most general measurement on the modes entering the photodetectos
of Fig.~\ref{fig:QKD-measurements}, namely, $A$, $B$, $C$, and
$D$, can be written in terms of the following measurement operators:
\begin{equation}
M_{abcd}=|a\left\rangle _{AA}\right\langle a|\otimes|b\left\rangle _{BB}\right\langle b|\otimes|c\left\rangle _{CC}\right\langle c|\otimes|d\left\rangle _{DD}\right\langle d|
\end{equation}
 for PNRDs, where $a,\, b,\, c,\, d=0,\,1$ and $|k\left\rangle _{K}\right.$
represents a Fock state for the optical mode $K=A,\, B,\, C,\, D.$
In the case of NRPDs, we only need to replace $|1\left\rangle _{KK}\left\langle 1|\right.\right.$
with $\left(I_{K}-|0\left\rangle _{KK}\left\langle 0|\right.\right.\right)$,
where $I_{K}$ is the identity operator for mode $K.$

Similarly, we can define the corresponding probabilities to the above
measurement operators as follows 
\begin{equation}
P_{abcd}=Tr\left(\rho_{ABCD}M_{abcd}\right).
\end{equation}
 The explicit forms for $P_{{\rm click}}$ and $P_{{\rm error}}$
are then given by 
\begin{equation}
P_{{\rm click}}=P_{C}+P_{E}\label{eq:equazione B4}
\end{equation}
 and 
\begin{equation}
P_{{\rm error}}=e_{d}P_{C}+(1-e_{d})P_{E},\label{eq:equazione B5}
\end{equation}
 where $e_{d}$ is the dephasing (misalignment) error, and 
\begin{equation}
P_{C}=\begin{cases}
(1-d_{c})^{2}(P_{1100}+P_{0011}+d_{c}(P_{1000}+P_{0100}+P_{0010}+P_{0001})+2d_{c}^{2}P_{0000}),\,\mbox{PNRD}\\
\left(\frac{d_{c}^{2}}{2}-d_{c}+1\right)(P_{1100}+P_{0011})+d_{c}(1-\frac{d_{c}}{2})(P_{1001}+P_{0110})\\
+\frac{d_{c}}{2}(2-d_{c})(P_{1000}+P_{0100}+P_{0010}+P_{0001})+\frac{d_{c}^{2}}{2}(2-d_{c})^{2}P_{0000}\\
+\frac{1}{2}(P_{1110}+P_{1101}+P_{0111}+P_{1011})+\frac{d_{c}}{2}(2-d_{c})(P_{1010}+P_{0101})+\frac{1}{2}P_{1111},\,\mbox{NRPD}
\end{cases}
\end{equation}
is the probability that Alice and Bob assign identical bits to their raw keys if there is no misalignment, and 
\begin{equation}
P_{E}=\begin{cases}
(1-d_{c})^{2}(P_{1001}+P_{0110}+d_{c}(P_{1000}+P_{0100}+P_{0010}+P_{0001})+2d_{c}^{2}P_{0000}),\,\mbox{PNRD}\\
\left(\frac{d_{c}^{2}}{2}-d_{c}+1\right)(P_{1001}+P_{0110})+\frac{d_{c}}{2}(2-d_{c})(P_{1000}+P_{0100}+P_{0010}+P_{0001})\\
+\frac{d_{c}^{2}}{2}(2-d_{c})^{2}P_{0000}+\frac{1}{2}(P_{1110}+P_{1101}+P_{0111}+P_{1011})\\
+\frac{d_{c}}{2}(2-d_{c})(P_{1100}+P_{1010}+P_{0011}+P_{0101})+\frac{1}{2}P_{1111},\,\mbox{NRPD}
\end{cases}
\end{equation}
is the probability that they make an erroneous bit assignment in the absence of misalignment. 

%

\end{document}